\documentclass[aps,prr,twocolumn,superscriptaddress]{revtex4-1}
\usepackage{amsfonts}
\usepackage{graphicx}
\usepackage{epstopdf}
\usepackage{epsfig}
\usepackage{amsmath}
\usepackage[normalem]{ulem}
\usepackage{multirow}
\usepackage{color}
\usepackage{makecell}
\usepackage{array}
\usepackage{booktabs}
\usepackage{mathtools}
\usepackage{bm}
\usepackage{color}
\usepackage{array}
\usepackage{booktabs}
\usepackage{tabularx}

\newcolumntype{Y}{>{\centering\arraybackslash}X}

\newcommand{\Rmnum}[1]{\expandafter\@slowromancap\romannumeral #1@}
\makeatother

\begin{document}

\title{Tomography of time-dependent quantum spin networks with machine learning}

\author{Chen-Di Han}
\affiliation{School of Electrical, Computer and Energy Engineering, Arizona State University, Tempe, Arizona 85287, USA}

\author{Bryan Glaz}
\affiliation{Vehicle Technology Directorate, CCDC Army Research Laboratory, 2800 Powder Mill Road, Adelphi, MD 20783-1138, USA}

\author{Mulugeta Haile}
\affiliation{Vehicle Technology Directorate, CCDC Army Research Laboratory, 2800 Powder Mill Road, Adelphi, MD 20783-1138, USA}

\author{Ying-Cheng Lai} \email{Ying-Cheng.Lai@asu.edu}
\affiliation{School of Electrical, Computer and Energy Engineering, Arizona State University, Tempe, Arizona 85287, USA}
\affiliation{Department of Physics, Arizona State University, Tempe, Arizona 85287, USA}

\begin{abstract}

Interacting spin networks are fundamental to quantum computing. Data-based tomography of time-independent spin networks has been achieved, but an open challenge is to ascertain the structures of time-dependent spin networks using time series measurements taken locally from a small subset of the spins. Physically, the dynamical evolution of a spin network under time-dependent driving or perturbation is described by the Heisenberg equation of motion. Motivated by this basic fact, we articulate a physics-enhanced machine learning framework whose core is Heisenberg neural networks. In particular, we develop a deep learning algorithm according to some physics motivated loss function based on the Heisenberg equation, which ``forces'' the neural network to follow the quantum evolution of the spin variables. We demonstrate that, from local measurements, not only the local Hamiltonian can be recovered but the Hamiltonian reflecting the interacting structure of the whole system can also be faithfully reconstructed. We test our Heisenberg neural machine on spin networks of a variety of structures. In the extreme case where measurements are taken from only one spin, the achieved tomography fidelity values can reach about $90\%$. The developed machine learning framework is applicable to any time-dependent systems whose quantum dynamical evolution is governed by the Heisenberg equation of motion.

\end{abstract}

\date{\today}

\maketitle

\section{Introduction} \label{sec:intro}

Quantum computation based on spin is a fundamental component of quantum
information science and technology~\cite{nielsen}. Recently, it has been
demonstrated that manipulating $50$ spins can generate the computational
capability beyond any kind of classical computers, leads to quantum
supremacy~\cite{neill2018blueprint,arute2019quantum}. From a network point
of view, the information exchange between any pair of spins can be regarded
as a link between the two spins. When the interactions associated with all
spin pairs are taken into account, the end result is effectively a network,
giving rise to the subfield called qubit or spin
networks~\cite{yurke1984quantum, cirac1997quantum,christandl2004perfect}.
Experimentally, a multispin coupling system can be realized using cavity
quantum electrodynamics~\cite{parkins1993synthesis}, ion
traps~\cite{duan2010colloquium}, or superconducting
qubits~\cite{neill2018blueprint,arute2019quantum}.

There are two types of spin networks: time-independent or time-dependent. In
the former case, the system can be decomposed into a sequence of quantum
gates~\cite{barenco1995elementary}, resembling a classical circuit structure.
Since the Hamiltonian is constant over time, this effectively leads to quantum
adiabatic computing systems, where quantum computing algorithms can be
performed on the ground states~\cite{aharonov2008adiabatic}. For
time-dependent spin networks, both the spin coupling and an external,
time-varying field~\cite{benjamin2003quantum} or an output control
signal~\cite{burgarth2010scalable} are present. It was argued that in
time-varying spin networks, the problem of switch off can be mitigated and
the computation speed can be enhanced~\cite{sjoqvist2012non}. In general, an
external field can serve to increase the computational capacity even for
relatively simple spin network structures~\cite{burgarth2010scalable,
glaser2015training}. However, it is challenging to analyze and realize
time-dependent control of spin networks. Recently, the idea of embedding a
time-dependent spin network into a time independent one was studied, but the
generality or universal applicability of this approach remains
unknown~\cite{banchi2016quantum,innocenti2020supervised}.

In recent years, the inverse problem of spin networks has attracted a great
deal of attention. The basic question is, given only limited access to the
system, i.e., only part of the system can be measured, can the global structure
of the spin network be determined? Previous efforts focused on monitoring the
Hamiltonian as a function of time through the Eigenstate Realization
Algorithm (ERA)~\cite{zhang2014quantum,zhang2015identification}, compressive
sensing~\cite{shabani2011estimation,magesan2013compressing}, or machine
learning~\cite{banchi2016quantum,innocenti2020supervised}. The basic idea is
to find the coefficients of the power series terms constituting the Hamiltonian
in some basis. However, when applying to time-dependent spin networks, these
approaches are limited to systems of a single spin or those with a special
type of external field~\cite{magesan2013compressing,de2016estimation,
bairey2019learning}. The general difficulty is that the functional form of the
time signal generates an optimization problem in infinite
dimensions~\cite{unger2019infinite}, rendering inapplicable any optimization
algorithm designed for finding a finite number of parameters. For the methods
based on the eigenstates, difficulties arise when the system changes too fast
with time~\cite{zhang2014quantum,zhang2015identification,bairey2019learning}.

In this paper, we solve a general class of inverse problems in spin
networks by exploiting machine learning~\cite{carleo2019machine}. Our work
was partly inspired by the recent work on the classical Hamiltonian Neural
Networks (HNNs)~\cite{de1993class,greydanus2019hamiltonian,
bertalan2019learning,choudhary2019physics}, where the basic idea is to
introduce a physics-based, customized loss function to ``force'' the dynamical
evolution of the system to follow that stipulated by the classical Hamilton's
equations. However, the existing HNNs are not directly applicable to quantum
spin networks, thereby requiring new approaches. Our idea originates from the
basic physical consideration that the dynamical evolution of quantum spin
networks is governed by the Heisenberg equations of motion. We are thus
motivated to develop a class of Heisenberg Neural Networks (HENNs) by
exploiting deep learning to predict the Hamiltonian but under the constraint
of the Heisenberg equations of motion. The HENNs have the advantage of
guaranteeing that the underlying quantum evolution possesses the Hermitian
structure. Our only assumption for the quantum spin networks is that their
Hamiltonian varies continuously with time, a situation that can be expected
to hold in experiments in general.

Our main results are the following. There are two types of networks involved
in our machine-learning framework for tackling the inverse problem: the
original time-dependent spin network whose structure is to be determined based
on incomplete local measurements and the HENN that is an artificial neural
network for predicting the Hamiltonian of the original system. We treat the
dynamical evolution of the original system in terms of both Schr\"{o}dinger and
Heisenberg pictures. We demonstrate that, with only local measurements, the
local Hamiltonian can be recovered, similar to the solution of the local
Hamiltonian learning problem~\cite{bairey2019learning}. In particular, defining
the tomography fidelity as the ratio between the correctly predicted links and
the total possible number of links in the underlying spin network, we find that
the fidelity can reach $90\%$ even when the number of spins measured is much
smaller than the system size. In fact, the predicted Hamiltonian contains the
global information about the coupling profile of the original spin network. We
note that the problem of network reconstruction or tomography has been well
studied in classical nonlinear dynamical systems~\cite{timme2007revealing,
shandilya2011inferring,su2012detecting,han2015robust,wang2016data}, and there
was also a study of structure identification for time-independent spin
networks~\cite{kato2014structure}. Our work goes beyond the relevant literature
in that we have successfully articulated and validated a general machine
learning framework of quantum tomography for time-dependent spin networks.

In Sec.~\ref{sec:system}, we describe the HENN learning framework. In
Sec.~\ref{sec:result}, we test our machine-learning method using a variety of
time-dependent spin networks, which include networks with short- or long-range
interactions and two quantum gates. In Appendix~\ref{Appendix_A}, we present
analytic results with HENNs for one- and three-spin systems.

\section{Time-dependent	quantum spin networks and Heisenberg neural networks} \label{sec:system}

Consider a system of spins coupled by an external field. The Hamiltonian is
\begin{equation} \label{eq:1_H}
H(t)=h^{(1)}+f(t)h^{(2)},
\end{equation}
where $h^{(1,2)}$ represent the time-independent Hamiltonian and $f(t)$ is a
continuous function of time that is the result of the application of a
time-dependent electrical or magnetic field. Suppose the system is initially
in the state $|\psi_0\rangle$ at $t=0$. In the Schr\"{o}dinger picture where
the state evolves with time but the operators are time-invariant, at time $t$
the expectation value of an operator $A$ is given by $\langle A \rangle_t$.
In the Heisenberg picture where the state does not change with time but
the operators do, an operator evolves according to the Heisenberg equation
\begin{equation} \label{eq:1_Heisenberg}
\frac{dA^H}{dt}=i[H^H(t),A^H(t)],
\end{equation}
where the superscript $H$ specifies that the corresponding matrix is in the
Heisenberg picture, $H^H(t)=U_{t,0}^\dagger H(t)U_{t,0}$, and $U_{t,0}$ and
$H(t)$ do not commute with each other due to the time dependence. Once
$H^H(t)$ is known, the corresponding Hamiltonian in the Schr\"{o}dinger
picture $H(t)$ can be determined. The goal is to solve the Heisenberg equation
based the observations of $A$.

Since Eq.~\eqref{eq:1_Heisenberg} is a set of linear equations in $H^H(t)$,
for any time $t$ the equations are solvable if the number of non-equivalent
equations is no less than the number of unknown elements. That is, the
non-commutative operators at all times, $A^H(t)$, are required to be known.
This is a key difference from time-independent systems, where $H^H(t)=H$ so
one operator at any time, $A^H(t)$, can be used as the non-commutative
operator. In this case, once the observations (e.g., time series) are
sufficient, the Hamiltonian can be fully determined~\cite{zhang2014quantum,
zhang2015identification}.

The number of independent elements in the Hamiltonian matrix provides another
angle to appreciate the complexity of the problem. In particular, for a system
with $n$ spins, at a specific time $t$, the Hamiltonian in Eq.~\eqref{eq:1_H}
can be represented by a Hermitian matrix in terms of the $N = 2^n$ linearly
independent states. There are altogether $N^2 = 4^{n}$ bases for an $N\times N$
Hamiltonian matrix that is Hermitian. To fully solve Eq.~\eqref{eq:1_Heisenberg}
will thus need all the $4^n$ measurements at a given time. For example, for a
two-spin system, there are four linearly independent states, so in principle 16
observations are needed. These observations can be generated by the direct
product of the Pauli matrices
$S_{\alpha,\beta}=\sigma_\alpha^1\otimes \sigma_\beta^2$,
where $\alpha$ and $\beta$ are integers ranging from $0$ to $3$, which
correspond to the identity and the three Pauli matrices
$\sigma_x,\sigma_y,\sigma_z$. For these $16$ matrices, one is an identity that
commutes with all other matrices. Consequently, we need at least $2^N-1=4^n-1$
measurements to fully determine the Hamiltonian.

When the quantum states of all spins can be measured, it is straightforward
to obtain the Hamiltonian matrix through Eq.~\eqref{eq:1_Heisenberg}. A
difficult situation is that only a small fraction of the spins in the network,
e.g., one or two, are externally accessible. Experimentally, measurements
or observations have been reported for one-, two-, and three-spin
systems~\cite{myerson2008high,chow2010detecting,neeley2010generation}.
The pertinent question is, what can we learn about the whole network system
when only local measurements in some subspace of the full space are available?
To address this question, we decompose the Hamiltonian as
\begin{equation} \label{eq:1_subspace}
H=\underbrace{H_{o}+H_i}_{H_{o'}}+H_h,
\end{equation}
where $H_o$ is the subspace Hamiltonian for the observed spins, $H_i$
represents the interaction between the observed and the inaccessible spins,
and $H_h$ is Hamiltonian for the inaccessible spins. Let $H_{o'}\equiv H_o+H_i$,
which is the sub-Hamiltonian that contains information directly related to
the observed spins.

\begin{figure} [ht!]
\centering
\includegraphics[width=\linewidth]{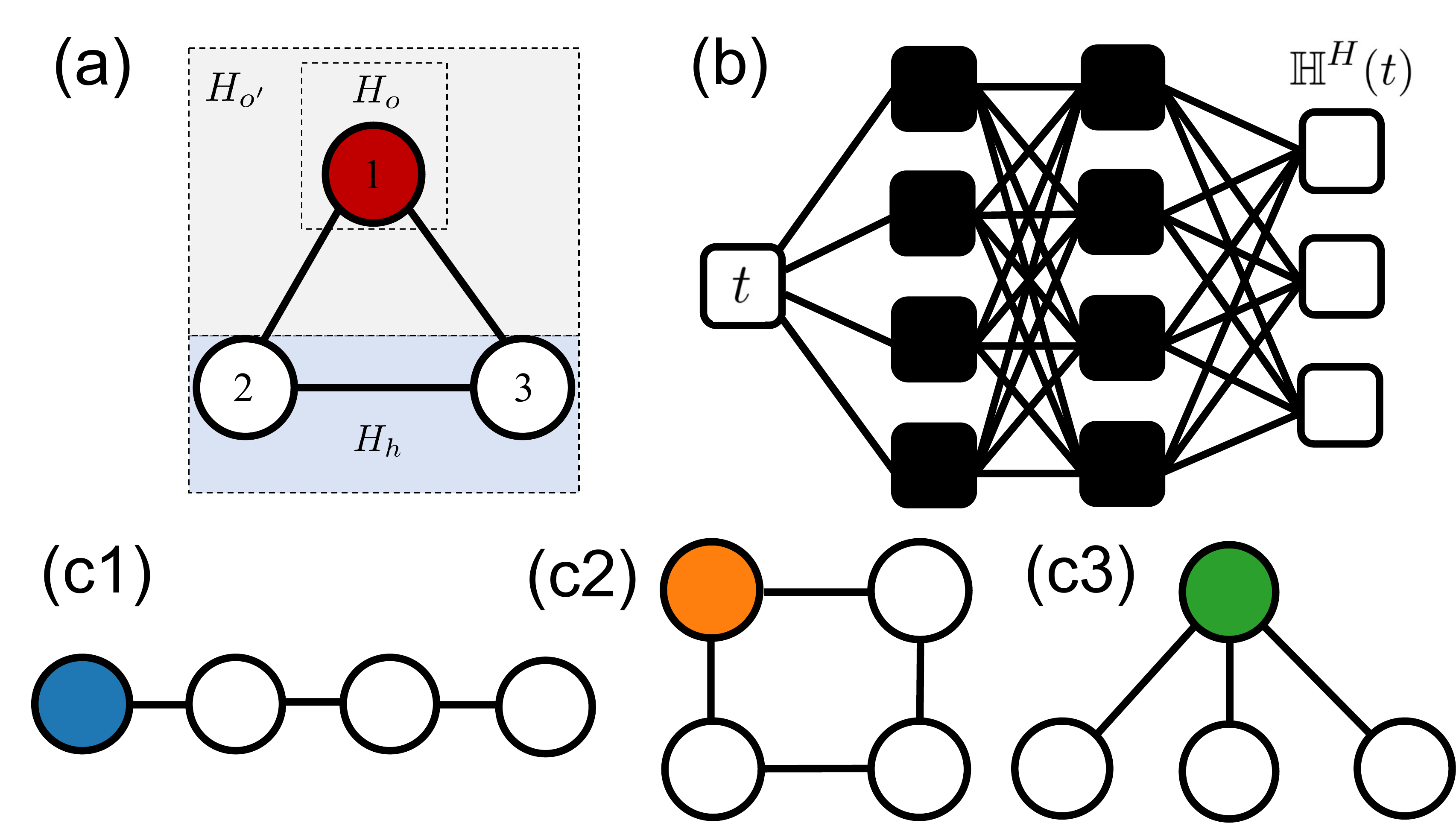}
\caption{Three- and four-spin networks and the machine learning architecture.
(a) Schematic illustration of local observation of the Hamiltonian in a
three-spin network. Say only one node or spin (the red one) can be observed,
which corresponds to the Hamiltonian $H_o$. The Hamiltonian for the hidden
nodes and their interactions are labeled as $H_{h}$. (b) Machine learning
(neural network) architecture, where the nodes in the hidden layers are
represented by black squares, the input and output are denoted by open
squares, and the various weighted links (solid line segments) connect the
input to the output. The input is one-dimensional: it is simply the time
variable $t$. The output constitutes the elements of the matrix
$\mathbb{H}^H(t)$ whose size is determined by the size of the spin network.
The weight associated with each link is calculated by the auto-gradient method
to minimize the custom loss Eq.~\eqref{eq:1_loss}. (c1-c3) Possible
architectures of a $4$-spin network, where the filled circle represents the
observable spin in the network. The networks in (c1-c3) have a chain, a cyclic,
and a tree structure, respectively. For $n=3$ and $n=5$, structures similar
to those in (c1-c3) exist.}
\label{fig:scheme}
\end{figure}

Take a three-spin system as an example, as shown in Fig.~\ref{fig:scheme}(a).
The three spins are labeled with $1,2,3$ and we assume that only the first
spin can be measured. The subspace of $H_{o}$ contains $3$ bases corresponding
to the Pauli matrices for the first spin $H_o=\sigma^1$, and $H_{i}$ contains
two-body interactions between the first spin and the second or the third spin
and the three-body interaction:
\begin{displaymath}
H_i=\sigma^1(\sigma^2+\sigma^3+\sigma^2\sigma^3).
\end{displaymath}
The subspace Hamiltonian $H_h$ contains the Pauli matrices for the second and
third spins as well as the two-body interaction between them:
$H_h=\sigma^2+\sigma^3+\sigma^2\sigma^3$. Overall, this is a three-node spin
network, where the nodal interactions represent different links. For each node,
three independent quantities (the three Pauli matrices) are needed to
characterize the spin polarization, which generate different combinations of
coupling. A unique feature of spin networks, which is not present in classical
complex networks, is that one link can couple more than two nodes.

The decomposition scheme in (\ref{eq:1_subspace}) is valid only in the
Schr\"{o}dinger picture. In the Heisenberg picture, different subspace are
mixed together in the time evolution, so all the subspace must be
simultaneously determined. For limited observations, the solutions of the
Heisenberg equation can be nonunique. To overcome this difficulty, we
exploit machine learning to predict the Hamiltonian. Inspired by the work
of HNN whose loss function is based on the Hamilton's equations of
motion~\cite{de1993class,greydanus2019hamiltonian,bertalan2019learning,
choudhary2019physics} for time-independent spin systems, we articulate
a general class of HENNs that conform with the Heisenberg equations of motion
with broad applicability to both time-dependent and time-independent spin
systems.

Figure~\ref{fig:scheme}(b) shows our neural network architecture with two
hidden layers. The input is one-dimensional: it is simply the time $t$. Each
layer is a convolution of the preceding layer:
$\mathbf{a}_l=\tau(\mathcal{W}_l \cdot \mathbf{a}_{l-1} + \mathbf{b}_l)$,
where $\mathbf{a}_l$ is the state vector of the $l$th layer, $\mathcal{W}$
is the weighted matrix connecting layers $l$ and $l-1$, $\mathbf{b}_l$ is
the bias vector of layer $l$, $\tau$ is a nonlinear activation function,
e.g., $\tau=\tanh$. The matrix $\mathcal{W}$ and the bias vector $\mathbf{b}_l$
are to be determined through training based on spin measurements. The output
is the Hamiltonian matrix in the Heisenberg picture. In our study, we use
two hidden layers, each with $200$ nodes. The neural network is built by
Tensorflow and the Keras package~\cite{chollet2015keras}. We use the Stochastic
Gradient Descent (SGD) and adaptive momentum (Adam) methods to determine the
optimal weighted matrix $\mathcal{W}_l$ and the bias vector $\mathbf{b}_l$ by
minimizing an appropriate loss function~\cite{kingma2014adam}. In particular,
we define our loss function as the mean square error in the time derivatives
of the observation as
\begin{equation} \label{eq:1_loss}
\mathcal{L}=\sum_{\text{Observations}}\left| \langle \dot{A}(t) \rangle_\text{real}- \langle \dot{A}(t) \rangle_\text{pred} \right|^2,
\end{equation}
where $\dot{A}_\text{pred}(t)=i[H^H(t), A(t)]$, and the matrix $H^H(t)$ is
the output of the HENN. Once the time derivatives for some given observations
are known, we input them to the loss function as the target to train the HENN
and subsequently to predict the Hamiltonian. Due to incomplete measurement and
finite optimization steps, the predicted Hamiltonian varies over different
rounds of training. It is thus necessary to take the statistical average of
the prediction and to calculate the variance.

It can be shown that the sub-Hamiltonian $H_{o'}$ in Eq.~\eqref{eq:1_subspace}
containing information directly related to the observed spins can be
recovered~\cite{bairey2019learning}. For the subsystem not directly related
to the observed spin, its Hamiltonian $H_h$ cannot be fully recovered.
However, we can show that, in the subspace of $H_h$, if the machine predicted
coupling value between two nodes is smaller than some threshold, then it
effectively indicates null coupling. This means that our HENN is capable
of determining the coupling configuration for the spin network based on
if the predicted Hamiltonian matrix elements are zero or finite, providing
a solution to the tomography problem for the whole system. In particular, the
tomography contains two types of information: whether the spins are coupled
and if so, how they are coupled. The first one is related to the spatial
structure of the network, as exemplified in Fig.~\ref{fig:scheme}(a), where
spin $1$ is coupled to spin $2$ and $3$. The second type of information gives
the the type of coupling among all possible coupling configurations determined
by the spin polarization vector at each node.

The prediction phase of our HENN thus consists of the following steps.

First, for a given quantum spin system, we take measurement $A$ from some part
of the system and calculate the corresponding matrix elements $\mathbb{A}^H(t)$
based on the linear equation
\begin{displaymath}
\langle \psi_0 | A^H(t)|\psi_0\rangle=\langle A\rangle_t.
\end{displaymath}
To obtain the matrix elements, the number of linearly independent initial
states must be larger than the number of independent elements of the matrix.
Specifically, for a spin network with $n$ spins, at least $4^n$ linearly
independent initial states are needed.

Second, we build up a neural network as in Fig.~\ref{fig:scheme}(b) with
input time $t$ and output as the matrix elements $\mathbb{H}^H(t)$. We train
the network using the loss function defined in Eq.~\eqref{eq:1_loss}. After
the HENN is properly trained, we evaluate the Hamiltonian for a given time
series, and convert it into the Schr\"{o}dinger picture. The coupling among
the nodes can be obtained from the decomposition
\begin{eqnarray}
	\nonumber
\mathbb{H}(t)&=& c_0(t)\mathbb{I}+\sum_{i,j} c_{ij}(t) \sigma^i_j, \\
	&+& \sum_{i,j,k,m} c_{ijkm}(t) \sigma^i_j \sigma^k_m+\cdots
	\equiv \sum_ic_i(t)S_i,
\end{eqnarray}
where $S_i$ is the basis of the $N$-dimensional Hamiltonian matrix and $c_i(t)$
is the corresponding coupling coefficients at time $t$. We choose $S_i$ to be
the direct product of the Pauli matrices plus the identity matrix. The
coefficients $c_i(t)$ determine the coupling configuration of the system.

Third, after obtaining the time series of the coupling coefficients, we take
the time average for each basis $\overline{c}_i=\int_0^t |c_i(t)|dt$ and
normalize them by their maximum value. We set some threshold: any value above
which indicates an existent coupling between the corresponding spins.

To better illustrate our HENN based machine learning procedure, in
Appendix~\ref{Appendix_A}, we present two explicit examples for HENN
predicted Hamiltonian: a one-spin system and a three-spin system.

\section{Results} \label{sec:result}

We test the predictive power of the proposed HENNs for a number of spin
systems. As noted, in a quantum spin network, the concept of links can be
quite different from those in classical networks. In particular, one link is
referred to as a specific way of coupling in the underlying spin network.
For a system with $n$ spins, the total number of linearly independent
states is $2^n$. The total number of independent elements in the
Hamiltonian matrix is $4^n$, which is the total number of possible
ways of coupling in the system. The links are generated by the direct
products of the Pauli and identity matrices. The types of links include
self-coupling, two-body interactions and long-range interactions. A quantity
to characterize the machine-learning performance is the tomography fidelity,
defined as the ratio between the number of correctly predicted links and the
total possible number of links. Disregarding the identity matrix, we define
the tomography fidelity as
\begin{equation} \label{eq:2_Fidelity_t1}
	F_t=\frac{4^n-1- (\text{\# of missing links})}{4^n-1},
\end{equation}
where the tomography is meaningful for $F_t>50\%$. A more useful characterizing
quantity is the success in identifying the structure of $H_h$, as this is proof
that the method can not only yield the structure of the subsystem from which
measurements are taken ($H_{o'}$), but also information about the complementary
subsystem from which no observations are made ($H_h$), so that information
about the whole system can be obtained. This alternative fidelity measure
is defined as
\begin{equation} \label{eq:2_Fidelity_t2}
F_{t'}=\frac{4^{n'}-1-(\text{\# of missing links})}{4^{n'}-1}.
\end{equation}
where $n' = n - n_{\text{obs}}$ and $n_{\text{obs}}$ is the number of spins
from which observations are taken.

\subsection{Tomography of spin networks based on two-body interactions}

The sub-Hamiltonians $h^{(1)}$ and $h^{(2)}$ in (\ref{eq:1_H}) of a spin
network with two-body interactions are given by
\begin{equation} \label{eq:2_two_body}
\begin{split}
h^{(1,2)}=&\sum_{i=1}^n \sum_{j=1}^3 c^{(1,2)}_{ij} \sigma^i_j+ \\
&\sum_{i=1}^n \sum_{j=i+1}^n \sum_{m=1}^3 \sum_{l=1}^3 w_{ij}c^{(1,2)}_{ijml} \sigma^i_m \sigma^j_l,
\end{split}
\end{equation}
where $c$'s are random numbers between $0$ and $1$. The superscript of $\sigma$
indicates the number of spins, which varies from $1$ to $n$, the
subscripts $1,2,3$ denote the $x$, $y$ and $z$ components of the spin,
respectively, $w_{ij}$ is the $ij$th element of the adjacency matrix as in
a conventional, undirected network, where $w_{ij} = 1$ indicates there is
coupling between spin $i$ and spin $j$, otherwise $w_{ij} = 0$. The first term
of $h$ contains self-couplings, and the second term contains two-body
couplings. Due to the exponential growth of the computational overload with
the number of spins in the network, we limit out study to networks with
$n \le 5$ spins. The Hamiltonian (\ref{eq:2_two_body}) arises in a variety of
physical situations such as the Heisenberg model or spin glass
systems~\cite{baxter,nishimori}. The time dependence in the general
Hamiltonian (\ref{eq:1_H}) is introduced into the network with the following
``driving'' function of time:
\begin{equation} \label{eq:2_ft}
f(t)=\sin(\omega t+2\pi\phi),
\end{equation}
where $\omega$ and $\phi$ are random numbers whose values are taken between
zero and one.

We test HENNs with the three structures shown in Figs.~\ref{fig:scheme}(a) and
\ref{fig:scheme}(c1-c3). The main difference among them lies in the degree of
the observed node. For example, for the networks in
Figs.~\ref{fig:scheme}(c1-c3), the degree of the observed spin is 9, 18, and
27, respectively. To generate the data, we choose $100$,
$300$ and $1,100$ random initial conditions for $n=3$, 4 and $5$. For
each initial condition, we numerically integrate the Heisenberg equation
(\ref{eq:1_Heisenberg}) for $0 < t < 5$, and extract from this time interval
100 equally spaced points as the measurement data. The calculated time series
for a given initial state correspond to observations of $\sigma_x$, $\sigma_y$
and $\sigma_z$ for the specific local spin in the network from which
measurements are taken. We take the time derivative defined in
Eq.~(\ref{eq:1_loss}) as the loss function for training the HENN. Following
the steps described in Sec.~\ref{sec:system}, we obtain the predicted
interaction structure of the network. Comparing with the actual structure
gives the tomography fidelity. Since the fidelity may vary for a different
Hamiltonian, for each specific type of networks, we repeat this process $100$
times.

\begin{figure} [ht!]
\centering
\includegraphics[width=\linewidth]{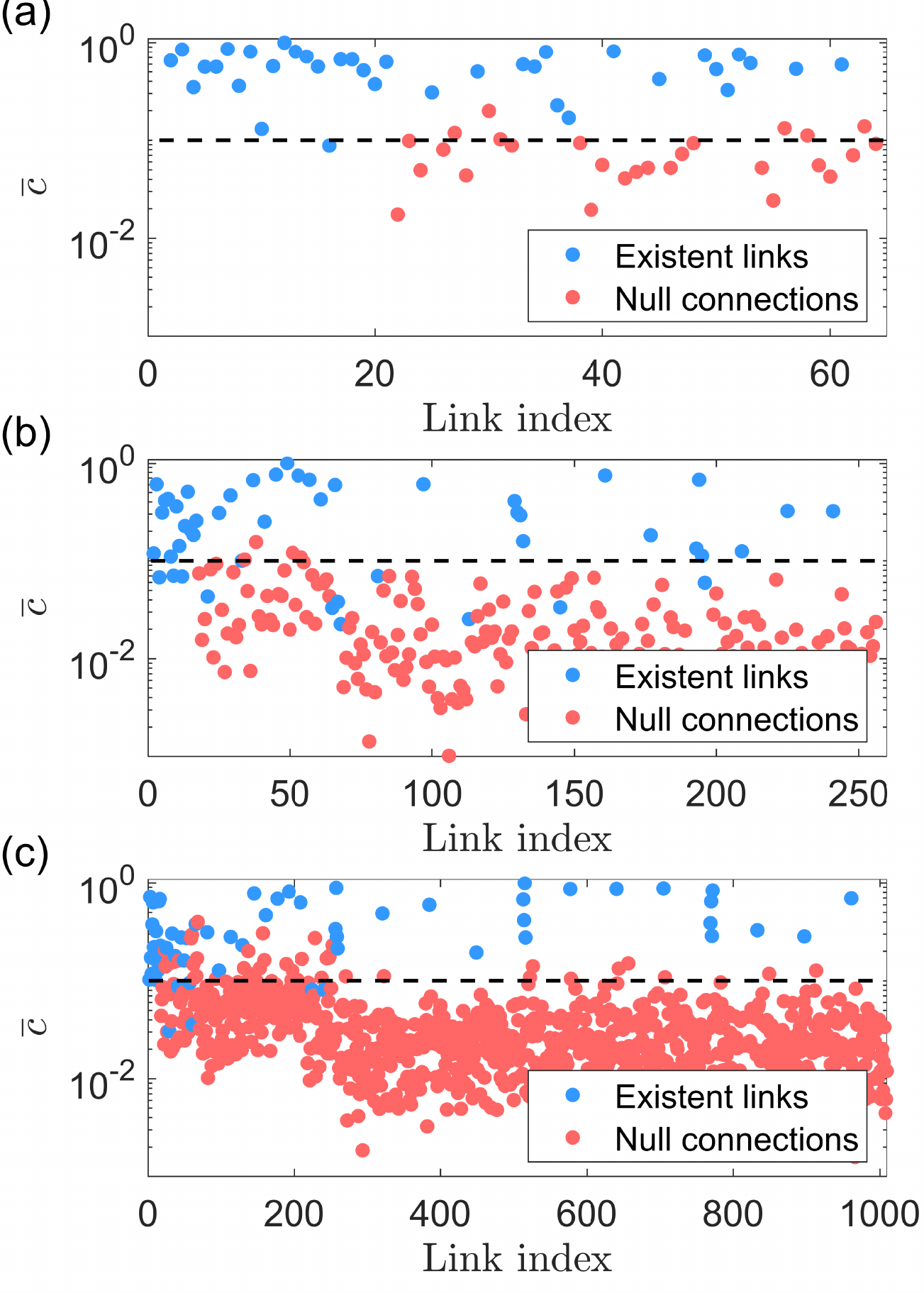}
\caption{ Tomography performance of HENN for cyclic networks of $n=3$, 4, and
5 spins. The network structure is given in Fig.~\ref{fig:scheme}(a).
(a) Reconstructed coupling value $\overline{c}$ versus the total number of
possible links. The result is for one of the spin networks with the average
tomography fidelity value from $100$ random realizations. The abscissa
represents the number of possible links for $n=3$, where the basis
ranges from $\sigma^1_0\otimes \sigma^2_0\otimes \sigma^3_1$ to
$\sigma^1_3\otimes\sigma^2_3\otimes\sigma^3_3$, where the total number of
possible links is $4^n-1=63$ (with the identity matrix taken away).
The blue dots represent the true, existent links, while nonexistent or null
links are denoted as the red dots. The horizontal dashed line is taken at
the $10\%$ of the predicted maximum coupling value. (b,c) Results from
$n=4$ and $5$, respectively, with the same legends. In all cases,
the horizontal dashed line can serve as a threshold for separating majority
of the existent links from majority of the null links, attesting to the
ability of the machine-learning scheme to infer the whole network structure
from local measurements only.}
\label{fig:3_cycle}
\end{figure}

Figure~\ref{fig:3_cycle}(a) shows the results of reconstructing the cyclic
network of three spins in Fig.~\ref{fig:scheme}(a), where the degree of each
node is 18 (excluding self-interactions) and there are 64 distinct links in
the network. What is displayed is the average predicted coupling value
$\overline{c}$ versus the link index, and the blue and red dots denote
the existent and null links, respectively. The dashed horizontal line defined
at $10\%$ of the maximum coupling value can separate majority of the existent
from majority of the null links. Figures~\ref{fig:3_cycle}(b) and
\ref{fig:3_cycle}(c) show the results from similar cyclic networks but with
four and five spins, respectively, with the same legends as those in
Fig.~\ref{fig:3_cycle}(a). These results indicate that, even when measurements
are taken from only one spin, the coupling structure of the time-dependent
Hamiltonian can be predicted by our HENN with a reasonably high accuracy.

A heuristic reason that the HENN is able to predict the structure of the
spin network correctly from only local measurements is as follows.
Recall that the input to the HENN is time, a continuous variable. The
differential property of the neural network guarantees that the predicted
Hamiltonian must be continuous and the time change for the predicted
Hamiltonian must follow the Heisenberg equation as stipulated by the physically
meaningful loss function. With these constrains, data from different time
will instill the correct physical relationships among the dynamical variables
into the neural network. As a result, the difficulty of non-uniqueness of
the solutions when solving the linear equations is overcome.

\begin{figure}
\centering
\includegraphics[width=\linewidth]{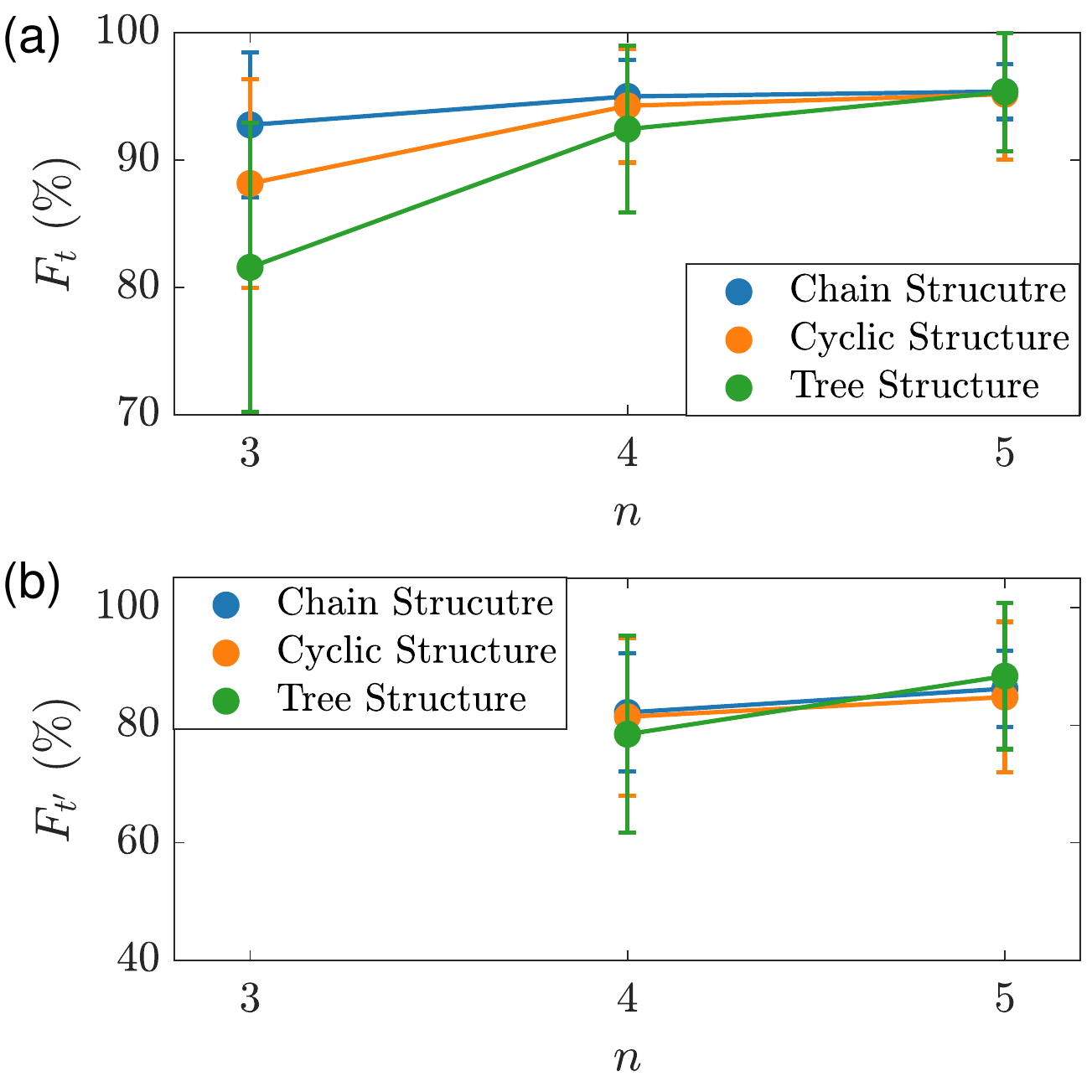}
\caption{Tomography fidelity measure. (a) Average tomography
fidelity $F_t$ and the standard deviation for nine distinct networks of three
types of structures in Figs.~\ref{fig:scheme}(c1-c3), respectively, each with
$n = 3$, 4, and 5 spins. In all cases, local measurements from only one
spin are taken. (b) Average alternative tomography fidelity $F_{t'}$ for
$n = 4$ and 5, corresponding to $n'= 3$ and 4, respectively.
(The results from $n = 3$ contain large statistical errors because of the
the relatively small size of the system and thus are not shown so that the
results for $n = 4$ and 5 can be seen clearly). In all cases, the fidelity
values are above $80\%$, indicating the predictive power of HENN.}
\label{fig:fidelity}
\end{figure}

To further characterize the performance of HENNs for different network
structures, we calculate the average fidelity measures for nine distinct
networks that include the chain, cyclic, and tree structures in
Figs.~\ref{fig:scheme}(c1-c3), respectively, each with $n = 3$, 4, and
5 spins, as shown in Figs.~\ref{fig:fidelity}(a) and \ref{fig:fidelity}(b)
for the measures $F_t$ and $F_{t'}$, respectively. For the network structures
in Figs.~\ref{fig:scheme}(c1-c3), the degrees of the measurement spin are 9,
18, and 27, respectively. For a fixed number of spins in the network, the
fidelity value decreases with the degree of the measurement spin. The reason
is that a larger degree of the measurement spin means that it is connected
with more ``hidden nodes'' in the network from which no measurements are
taken. As a result, the predicted coupling values are close to each other and
it becomes increasingly more difficult to distinguish them. Note that the
measure $F_t$ is defined for the whole network, which takes into account
not only the links between the measurement spin and the hidden spins, but also
the links among the hidden spins, where the latter are characterized by the
alternative tomography fidelity measure $F_{t'}$. Since this measure is purely
for the hidden spins from which no measurements are taken, we expect its value
to be lower than that of $F_t$, as shown in Fig.~\ref{fig:fidelity}(b).
In spite of the reduction in comparison with $F_t$, we see that the values
of $F_{t'}$ for $n = 4$ and $n = 5$ are still relatively high: approximately
$80\%$ and larger, attesting to the power of our HENN scheme to extract
information from the hidden spins.

Utilizing two-body coupling spin networks to evaluate the performance of
the HENNs has certain limitations. In particular, for a given network
structure, when the number of spins increases, the error appears to decrease,
due mostly to the exponential growth in the total possible number of links
in the network, which is an artifact. In some cases, the prediction results
can be trivial as the system size increases. For example, if all the two-body
couplings are null, then for $n=3$, the tomography fidelity value will be
about $40\%$ because approximately $60\%$ of the links are of the two-body
type. Similarly, for $n=4$ and 5, approximately $20\%$ and $5\%$ of the links
are of the two-body type, leading to artificial fidelity values of about
$80\%$ and $95\%$, respectively. Comparing with the results in
Fig.~\ref{fig:fidelity}, for $n=3$ and 4, the trivial prediction gives lower
fidelity values, but the difference diminishes for $n=5$. Consequently,
based solely on two-body interactions, that the tomography fidelity increases
with the system size is not synonymous to a better performance of the algorithm
for larger systems. For accurate tomography of quantum spin networks,
long-range interactions must be included.

\subsection{Tomography of quantum spin networks with long-range interactions}

We consider the more general Hamiltonian that contains all short- and
long-range interactions. Physical applications include the development of
quantum gates such as the Toffoli or the Fredkin gate that requires three-body
interaction~\cite{fredkin1982conservative}, spin glass with infinite-range
interactions~\cite{nishimori}, and quantum computing that requires high
coherence~\cite{neeley2010generation,dicarlo2010preparation}. We decompose
the Hamiltonian into two components, $h^{(1,2)}$, as in Eq.\eqref{eq:1_H},
which are given by
\begin{equation} \label{eq:2_random}
h^{(1,2)}=\sum_{i_1,i_2,\cdots, i_n =0}^3 r c^{(1,2)}_{i_1i_2\cdots i_n} \sigma^1_{i_1}\sigma^2_{i_2} \cdots \sigma^n_{i_n},
\end{equation}
where $c^{(1,2)}_{i_1i_2\cdots i_n}$ are random numbers between $0$
and $1$, and $r$ takes on the values of one or zero with equal probabilities.
The network comes into existence only for $r=1$. The function $f(t)$ rendering
the system time-dependent is chosen according to Eq.~\eqref{eq:2_ft}.

We consider systems with $n=3$, 4 or $5$ spins with $100$, $300$ and
$1100$ random initial conditions, respectively. Observing one spin leads
to time series of $\sigma_x$, $\sigma_y$ and $\sigma_z$ from this spin.
If two spins can be measured, we choose the observation variable to be
$\sigma_\alpha^1\otimes \sigma_\beta^2$, where $\alpha$ and $\beta$ are
integers from zero to three, corresponding to the identity and the three
Pauli matrices, respectively. Excluding the identity operation, we have
15 measured time series of 100 equally spacing points in the time interval
$0 < t < 5$.

\begin{figure} [ht!]
\centering
\includegraphics[width=\linewidth]{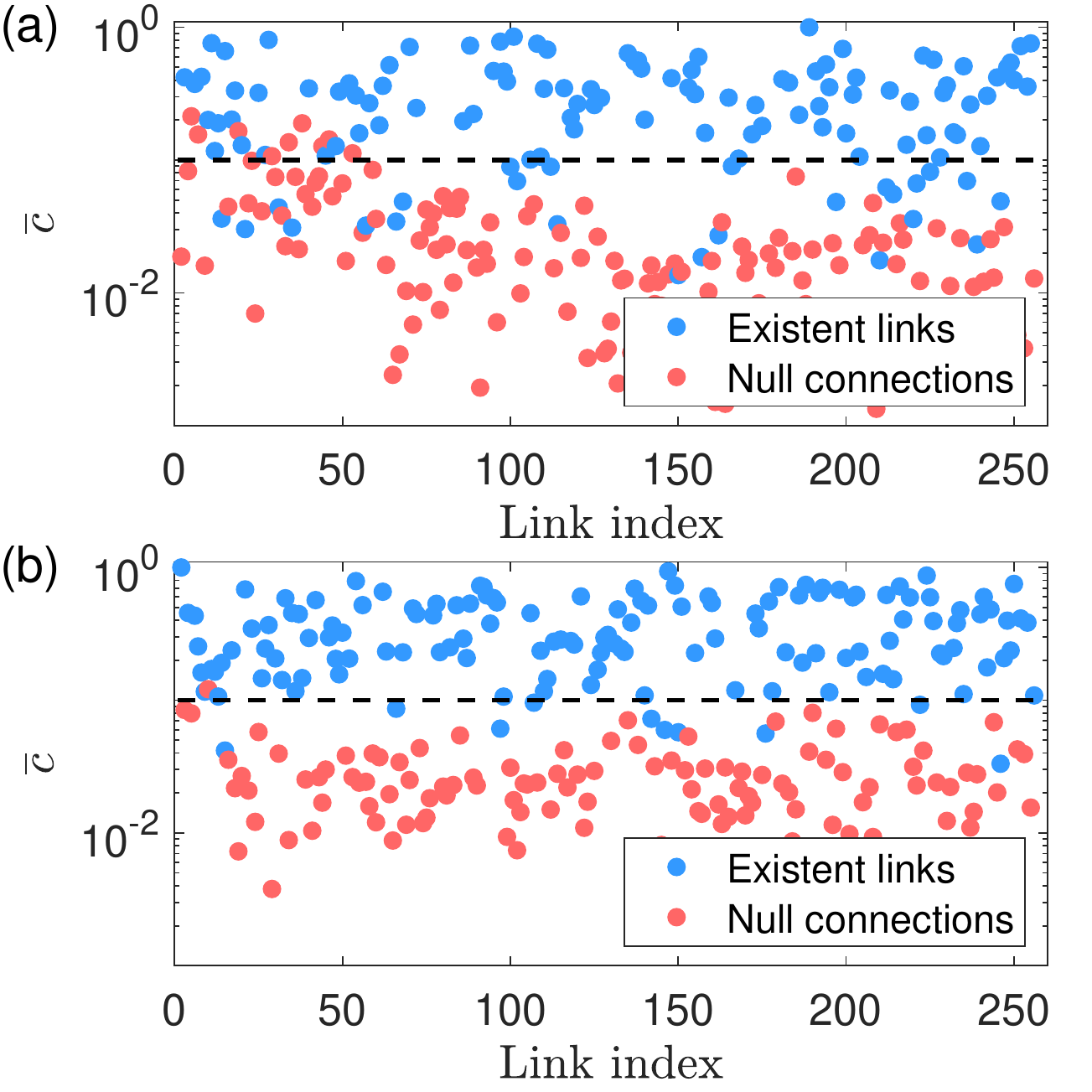}
\caption{ Prediction performance of HENN for a network of four spins with
long-range interactions. (a) Reconstructed coupling coefficients versus
based on time series measured from one spin only. The results are from one
realization of the spin network with the fidelity value equal to the average
fidelity value over 100 random realizations. The legends are the same as
those in Fig.~\ref{fig:3_cycle}. (b) The corresponding results when two spins
are observed with the time series as described in the text.}
\label{fig:LRSN_4}
\end{figure}

Following the procedure described in Sec.~\ref{sec:system}, we train the
HENN to predict the coupling configurations of the spin networks with
long-range interactions. Unlike the case where only two-body interactions
are taken into account, here the links are chosen randomly: we consider all
possible links and any specific link exists or does not exist with equal
probabilities. Figure~\ref{fig:LRSN_4} shows the prediction performance for
a network of $n=4$ spins, where panels (a) and (b) correspond to the
cases of measuring one and two spins, respectively. When only one spin is
measured [Fig.~\ref{fig:LRSN_4}(a)], most of the existent and nonexistent
links can be distinguished by the $10\%$ threshold line, yet there are still
quite a few links that are on the ``wrong'' side. When two spins are measured,
the prediction accuracy is higher as there are far fewer incorrectly predicted
links. This is intuitively reasonable as measuring more spins is equivalent
to imposing more constraints on the predicted Hamiltonian so as to improve
the prediction accuracy.

\begin{figure} [ht!]
\centering
\includegraphics[width=\linewidth]{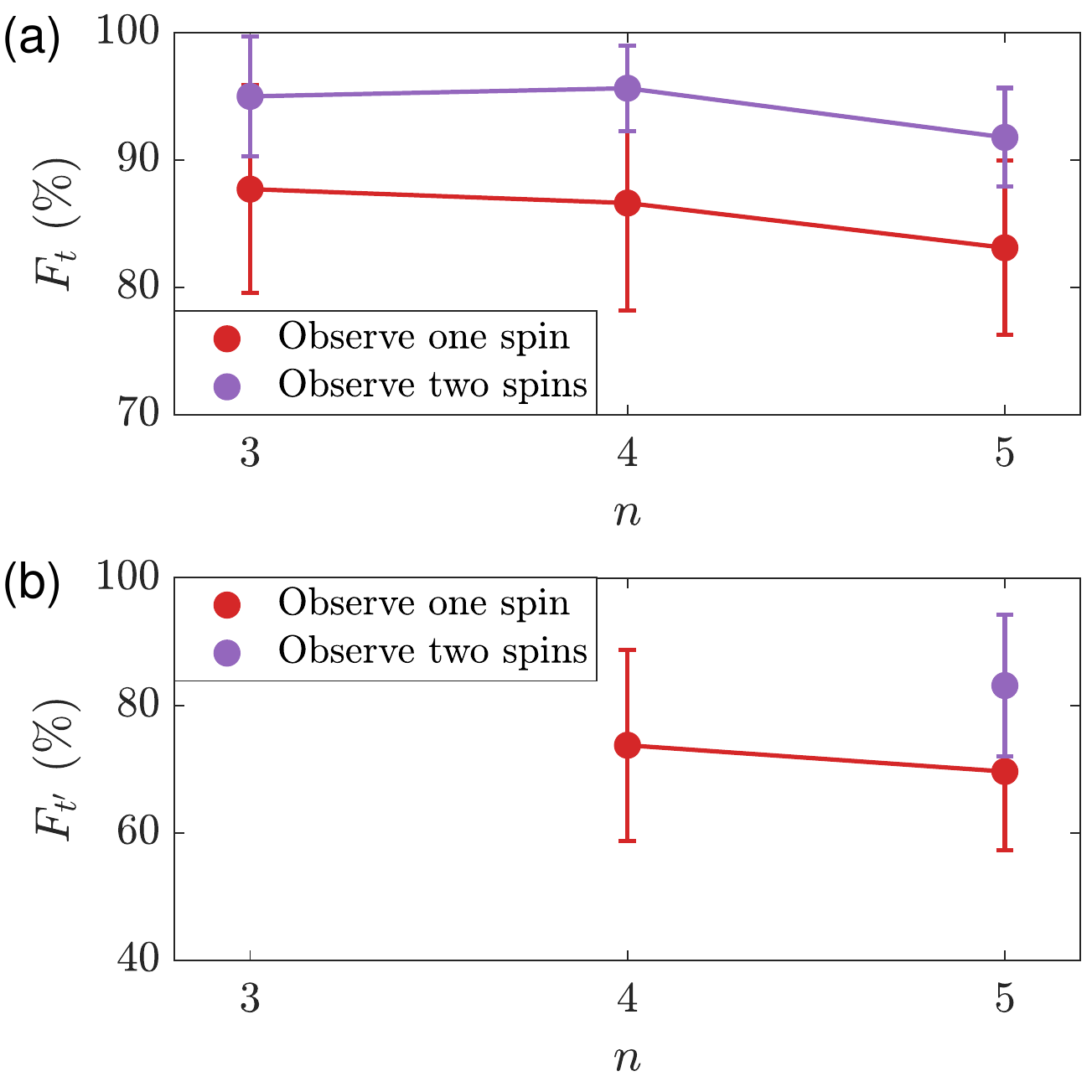}
\caption{ Fidelity of predicting random spin networks from observing one spin
or two spins. (a) Average tomography fidelity $F_t$ and the standard deviation.
Each data point is the result of averaging over 100 random initial-condition
realizations. (b) The corresponding results for $F_{t'}$. In the case of
observing one spin, results for $n = 3$ contain large fluctuations. When
two spins are observed, only the five-spin network generates reasonable values
of $F_{t'}$. In general, observing two spins leads to higher fidelity values.}
\label{fig:LRSN_5}
\end{figure}

Figure~\ref{fig:LRSN_5} shows the fidelity measure of predicting random
networks of $n = 3$, 4, and 5 spins, from observing one spin or two
spins. As shown in Fig.~\ref{fig:LRSN_5}(a), the fidelity value decreases
as the number of spins increases. This is expected because, when observing a
fixed number of spins, a larger system means more hidden spins and leads to
larger prediction uncertainties. Another expected feature is that, for
a fixed system size, observing two spins leads to higher fidelity values
[about $95\%$ in Fig.~\ref{fig:LRSN_5}(a)] as compared with the case of
observing one spin [about $85\%$ in Fig.~\ref{fig:LRSN_5}(a)].
Figure~\ref{fig:LRSN_5}(b) shows that the fidelity measure with respect to
the hidden structure exceeds $50\%$, indicating that the interactions among
the hidden spins can be predicted with statistical confidence. In fact, as
stipulated by Eq.~\eqref{eq:2_random}, our HENN can predict not only the
existence of the interactions but also their strength as characterized by
the coefficients $c^{(1,2)}_{i_1i_2\cdots i_n}$.

\subsection{Tomography of quantum gates}

We apply our HENN framework to a class of systems that are fundamental to
quantum computing: quantum logic gates. Such a gate typically consists of
two or three coupled spins~\cite{nielsen}. To be concrete, we consider the
Toffoli and Fredkin gates with three spins~\cite{fredkin1982conservative}
and demonstrate that HENN can perform the tomography. Experimentally, these
quantum logic gates can be implemented with optical
devices~\cite{lanyon2009simplifying,patel2016quantum} or superconducting
qubits~\cite{fedorov2012implementation}.

Toffoli gate is a Control-Control Not gate, i.e., when the first and
second spins have the signal $|1 1 \rangle$, the third spin will
flip~\cite{fredkin1982conservative}, which requires certain time, e.g.,
$t=1$. By this time, the evolution operator is
\begin{equation}
U_\text{Toffoli}=\begin{pmatrix}
\mathbb{I}_6 & \\
& \mathbb{X}(t=1)
\end{pmatrix},
\end{equation}
where $\mathbb{I}_6$ is the $6\times 6$ identity matrix,
\begin{equation}
\label{eq:2_X}
\mathbb{X}(t=1) =\begin{pmatrix}
0 & 1 \\
1 &0
\end{pmatrix}
\end{equation}
flips the third spin, and the off-diagonal blocks are zero. Similarly, the
time evolution operator for the Fredkin gate is
\begin{equation}
U_\text{Fredkin}=\begin{pmatrix}
\mathbb{I}_5 & & \\
& \mathbb{X}(t=1) & \\
&& 1
\end{pmatrix}.
\end{equation}
A physical constraint for $\mathbb{X}(t)$ is that, at $t=0$, the system does
not evolve, so $\mathbb{X}(t=0)=\mathbb{I}_2$. To build such a time evolution
operator, one can use the underlying time-independent Hamiltonian as a base
(Appendix~\ref{Appendix_B}), where the elements of $\mathbb{X}(t)$ are periodic
functions with the fundamental frequency as the flipping rate. Searching for
possible forms of $\mathbb{X}(t)$ is a basic issue in designing quantum logic
gates~\cite{banchi2016quantum, innocenti2020supervised}.

To demonstrate the applicability of HENN to quantum logic gates in a concrete
manner, we choose $\mathbb{X}(t)$ as
\begin{equation}
\mathbb{X}(t) =\frac{1}{2}\begin{pmatrix}
1+\exp(i\pi t) & 1-\exp(3i\pi t) \\
1-\exp(3i\pi t) & 1+\exp(i\pi t)
\end{pmatrix}
\end{equation}
to generate the time-dependent Hamiltonian. For training, we generate time
series from $t=0$ to $t=1$ with the time step $dt=0.01$ from $100$ random
initial conditions, and the time evolution of the dynamical variables of the
third spin is taken as the measurements. For comparison, we calculate the
tomography fidelity for both time-dependent and the corresponding
time-independent systems.

\begin{table} [ht!]
\caption{Tomography fidelity for Toffoli and Fredkin gates}
\label{ta:Quantum_Gate}
\begin{tabularx}{\linewidth}{YYY}
\hline\hline
\specialrule{0em}{1pt}{1pt}
& Toffoli gate & Fredkin gate \\
\specialrule{0em}{1pt}{1pt}
\hline
\specialrule{0em}{1pt}{1pt}
Time-dependent &  $81\%  \pm 3\%$ &  $87\%  \pm 3\%$ \\
\specialrule{0em}{1pt}{1pt}
Time-independent & $85\%  \pm 4\%$  &  $92\%  \pm 3\%$ \\
\specialrule{0em}{1pt}{1pt}
\hline\hline
\end{tabularx}
\end{table}

Table~\ref{ta:Quantum_Gate} lists the values of the tomography fidelity for
both Toffoli and Fredkin gates. It can be seen that the average tomography
fidelity for the Fredkin gate is larger than that for the Toffoli gate. Both
gates have seven links, but the Fredkin gate has more three-body coupling
terms than the Toffoli gate. This means that, for the Fredkin gate, more links
are directly connected to the observed spin. When the system becomes
time-independent, the fidelity values are slightly higher.

\subsection{Quantum tomography under noise}

Noise arising from random coupling with the environment will impact the quality of quantum tomography. Previously, the issue of noise was studied in the
context of Hamiltonian learning~\cite{zhang2014quantum,bairey2019learning}.
Here, we study the effect of noise on our HENN based quantum tomography.
A relevant point is that HENN can predict the correct form but only of the
local Hamiltonian, so the Hamiltonian for the whole spin system cannot be
uniquely determined. This is similar to obtaining the local Hamiltonian
from local observations~\cite{bairey2019learning}. To study the effect of
noise on quantum tomography, it is thus useful to compare the fidelity $F_t$
with the local Hamiltonian fidelity $F_{o'}$ defined as
\begin{equation} \label{eq:2_Fidelity_o2}
	F_{o'}=1-\frac{\| H_{{o'} (\text{pred})} - H_{{o'} (\text{real})} \|}{\| H_{{o'} (\text{real})} \|},
\end{equation}
where $H_{{o'}(\text{real})}$ and $H_{{o'}(\text{pred})}$ are the real and
predicted local Hamiltonian, respectively.

As a concrete example, we study a cyclic network of four spins, where
measurements are taken from only one spin, which are subject to additive
white Gaussian noise. The local Hamiltonian $H_o$ constitutes the three
Pauli matrices for the observed spin, giving rise to three measured time
series. The Hamiltonian $H_{o'}$ consists of the three Pauli matrices for
the observed spin and all the interactions that involve the observed spin.
In particular, there are $27$ two-body interactions, $81$ three-body
interactions and $81$ four-body interactions. As a result, $H_{o'}$
can generate a total of $192$ time series. From a different perspective, for
a four spin system, the total number of dynamical variables is 256. When
we measure one spin so that the other three spins are hidden, $H_h$ contains
one fourth of the dynamical variables ($4^3$) while $H_{o'}$ has the remaining.

\begin{figure} [ht!]
\centering
\includegraphics[width=\linewidth]{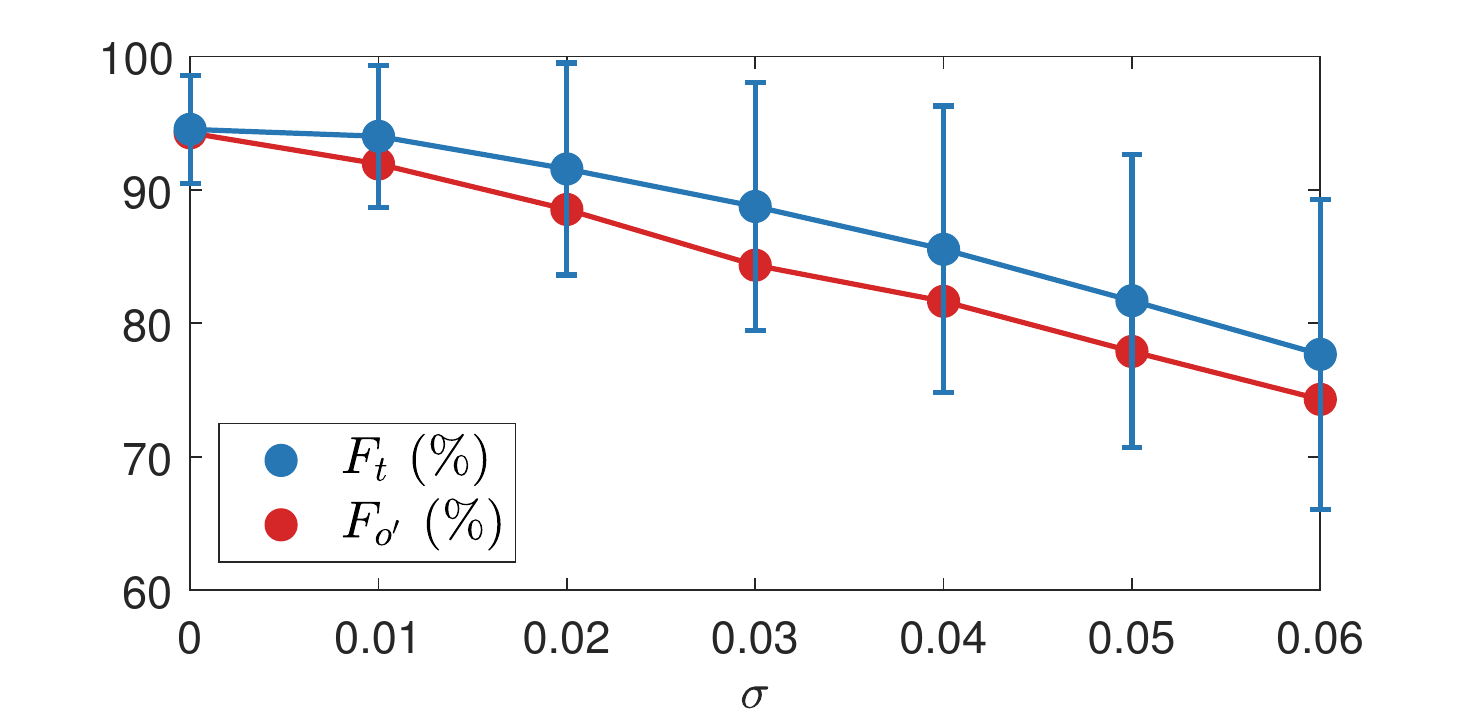}
\caption{ Network tomography fidelity under Gaussian noise. The system is a
cyclic network of four spins constructed according to Eq.~\eqref{eq:2_two_body},
where measurements are taken from one node. The additive white Gaussian noise
has the variance $\sigma=0.06$, which is approximately one third of the
average variation of the measured time series. The two sets of data
represent the two fidelity measures $F_t$ and $F_{o'}$ versus the noise
amplitude, where $F_{o'}$ characterizes the difference between the recovered
local Hamiltonian and the true one.}
\label{fig:noise}
\end{figure}

Figure~\ref{fig:noise} shows that the fidelity value decreases, albeit slowly,
as the noise amplitude increases. The robustness of HENN against weak noise
roots in the goal of HENN: finding one Hamiltonian that minimizes the loss
function. However, for strong noise, the derivatives in the Heisenberg
equation will generate unstable solutions. Figure~\ref{fig:noise} also reveals
the similarity between local Hamiltonian recovery and the tomography of the
whole system. In particular, it demonstrates that HENN can recover not only
the local Hamiltonian, but also the hidden structure of the spin network
which is not a simple extension of the local Hamiltonian.

\section{Discussion} \label{sec:discussion}

In quantum tomography, learning time-dependent systems from partial and
limited measurements remains a challenge, as it requires optimization in an
infinite-dimensional space. Machine learning provides a potentially viable
solution. Because of the underlying physics of the spin systems, it is
necessary to incorporate the physical constraints into the learning
algorithms. Historically, the idea of developing physics informed artificial
neural networks was conceived almost three decades ago~\cite{de1993class}, but
it has recently attracted a revived interest, particularly in the context of
HNNs~\cite{greydanus2019hamiltonian,bertalan2019learning,choudhary2019physics}.
Our idea is that, for time-dependent quantum systems, the Heisenberg
representation is natural in which the operators evolve with time as governed
by the Heisenberg equation. Neural networks taking into account the physical
constraints manifested as the Heisenberg equation may thus provide an approach
to tomography of time-dependent quantum systems. Using spin networks that
have been exploited extensively in quantum computing as a paradigm, we have
developed a class of Heisenberg Neural Networks (HENNs) and demonstrated that,
based on the time series measurements of local spin variables, not only the
local Hamiltonian but that of the whole spin system can be faithfully
determined. The method is effective even when measurements are conducted on a
small part of the system, e.g., measuring one spin in a five-spin system.
Considering that the existing algorithms on quantum tomography of spin
networks were designed for networks whose structures are completely
known~\cite{burgarth2009coupling,di2009hamiltonian,lapasar2012estimation},
our work represents a useful complement.

For quantum tomography of time-dependent interacting spin systems, we have
tested a variety of network structures. In general, the tomography fidelity
depends on the interacting structure of the network. For example, it is
inversely proportional to the degree of the spin from which measurements are
taken (Fig.~\ref{fig:fidelity}) when the spin network is relatively dense, as 
a large degree means more interactions with the hidden spins that are not 
under observation. The fidelity value also depends on the number of observed 
spins relative to the total number of spins in the network where, naturally, 
measuring more spins can lead to higher fidelity values (e.g., 
Fig.~\ref{fig:LRSN_5}). Indeed, a comparison of the tomography results from 
the Toffoli gate with those from the Fredkin gate reveals explicitly that more 
coupling links with the spin being measured lead to increased fidelity values, 
when the spin network is relatively sparse.

A number of factors can affect the tomography accuracy.
For the spin networks studied, the choices of the coupling and the time
dependence as characterized by the driving function $f(t)$ are arbitrary.
Our study has revealed that the tomography quality does not depend on the
time signal insofar as the length of the measured time series is proper,
but the HENN predicted Hamiltonian tends to deviate from the true one after
approximately half of the driving period. If the time series are too short,
e.g., a fraction of the driving period, or if the time series are too long,
e.g., more than a few driving periods, the resulting fidelity value would
decrease. Another factor that can affect the fidelity is heterogeneity in the
couplings in the network. For the results in this paper, the distribution of
the couplings in the spin network is assumed to be uniform, where the typical
fidelity value achieved is about $90\%$. However, we find that large
variations in the coupling strengths can make the HENN ineffective. The ratio
between self- and mutual couplings can also affect the tomography, where if
the former dominate the latter, the errors in the tomography can be reduced.

Possible extensions of this work are as follows. When implementing the HENN,
the initial states must be specified with a number of constraints. That is, it
is necessary to know the initial quantum state of each spin initially. For
time-independent systems, quantum tomography of spin networks is possible even
if the initial states are not completely specified~\cite{di2009hamiltonian}
or if the dimension of the system with a given coupling structure needs to
be determined~\cite{sone2017exact,haehne2019detecting}. It would be useful
to study if these approaches can be extended to time-dependent systems with
partial initial conditions or incomplete information about the system.
Another issue is scalability. While our HENN is a type of deep learning
neural networks, the number of parameters to be optimized cannot be too large
due to computational overload. Recently, neural networks with more than $100$
billions parameters have been constructed~\cite{shazeer2017outrageously}, with
applications to large systems with 100 spins~\cite{bairey2019learning}.
To scale our HENN algorithm to large time-dependent spin systems remains to
be a challenge. The third issue is noise, which is of paramount importance
in quantum computing applications. For the spin network, we have obtained
preliminary results revealing that noise in the input time series can induce
instability in the time derivatives. To develop an effective noise-reduction
scheme (e.g., through proper filtering) to stabilize the time derivatives in
the HENN is an open problem.

Recently, there have been efforts in understanding machine learning (the field
of explainable machine learning), and physics-enhanced machine learning
represents a useful perspective~\cite{iten2020discovering}. The HENN
articulated in this paper, which includes time correlation, provides an
effective way to ascertain and understand the hidden structures in the neural
network. As such, our HENN may be exploited as a paradigm for explainable
machine learning.

\section*{Acknowledgement}

This work was supported by Army Research Office through Grant 
No.~W911NF-17-S-002.

\appendix

\section{Examples of Heisenberg neural networks} \label{Appendix_A}

We present a number of examples for which the HENN can be explicitly
constructed.

\subsection{One Spin System}

The Hamiltonian of a single spin system under a periodic driving is
\begin{equation} \label{eq:A1_Hamltonian}
H(t)=\sigma_x\sin(t).
\end{equation}
We use the convention $\sigma_x=\sigma_++\sigma_-$ and
$\sigma_y=i(-\sigma_++\sigma_-)$, where $\sigma_{+,-}$ correspond to the
creation and annihilation operators, respectively. In the basis $|1\rangle$
and $|0\rangle$, the matrix expressions of $\sigma_x$ and $\sigma_y$ are
\begin{equation}
\sigma_x=\begin{pmatrix}
0 & 1 \\ 1 & 0
\end{pmatrix} \ \mbox{and} \
\sigma_y=\begin{pmatrix}
0 & -i \\ i & 0
\end{pmatrix}.
\end{equation}
The third spin operator is defined by $\sigma_z|1\rangle=|1\rangle$ and
$\sigma_z|0\rangle=-|0\rangle$.

The HENN for the one-spin system can be constructed by following the three
steps described in Sec.~\ref{sec:system}, as follows.

\paragraph*{Step 1:} We generate random initial conditions $|\psi_0\rangle$
given by
\begin{equation} \label{eq:A1_initial}
\psi_0 =\frac{1}{r_1+r_2}\binom{\sqrt{r_1}\exp(i2\pi\theta_1)}{\sqrt{r_2}\exp(i2\pi\theta_2)}=\binom{\phi_{11}+i\phi_{12}}{\phi_{21}+i\phi_{22}},
\end{equation}
where $r_{1,2}$ are the initial probabilities in the respective state,
$\theta_{1,2}$ are the corresponding phase variables, both $r_{1,2}$ and
$\theta_{1,2}$ are uniform random numbers between zero and one. In the
machine learning algorithm, all quantities are real, so it is necessary to
convert the wavefunction into the summation for the real and imaginary parts.

\begin{figure} [ht!]
\centering
\includegraphics[width=\linewidth]{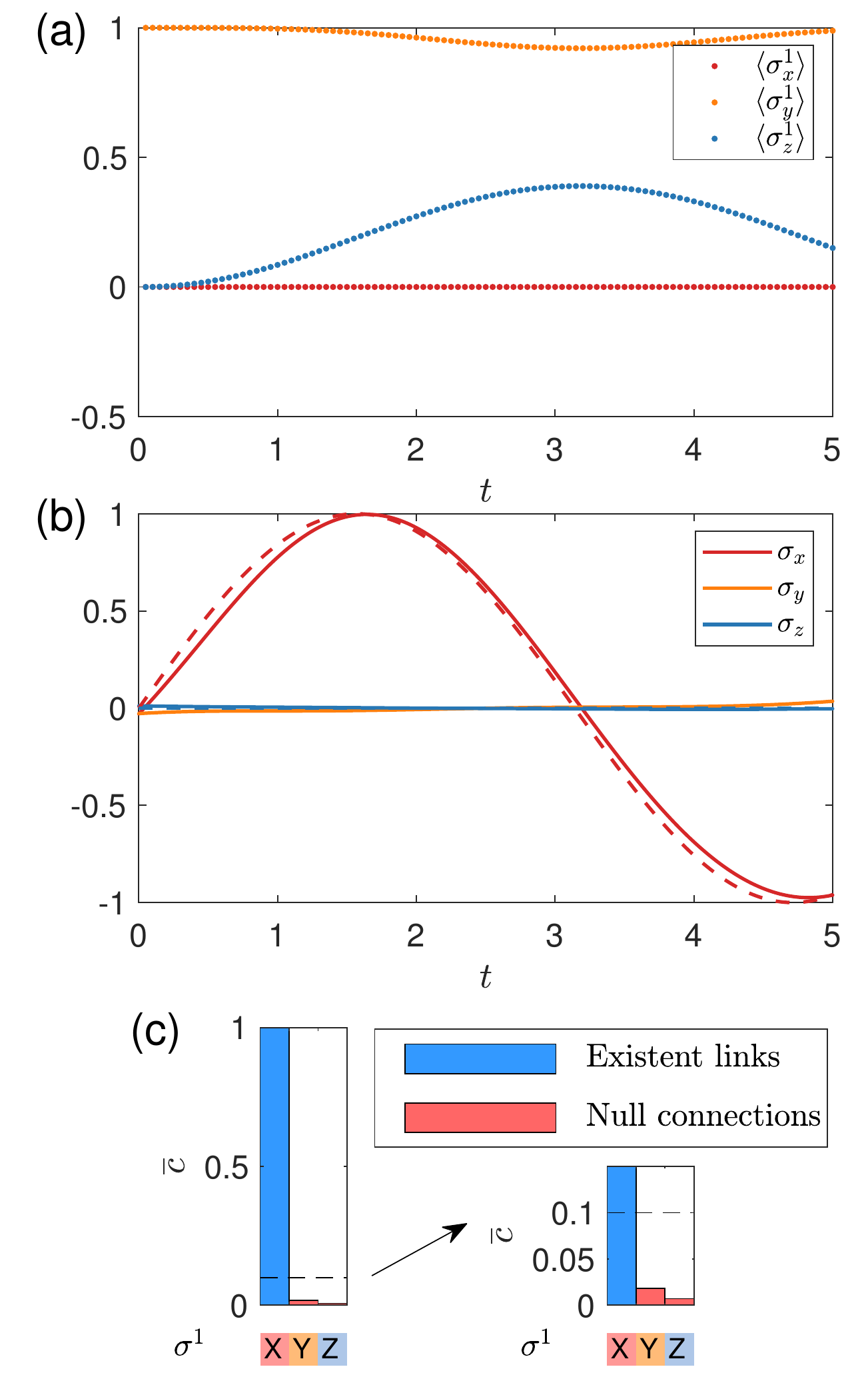}
\caption{ HENN based tomography of a single-spin sysem. (a) Time series of
$\sigma_x$, $\sigma_y$ and $\sigma_z$, where the initial state is
$(|1\rangle+i|0\rangle)/\sqrt{2}$. The dots correspond to the sampled
measurements. (b) The predicted Hamiltonian in the Schr\"{o}dinger picture,
where the three curves correspond to the decomposition in the three base
states and the dashed curves are the true values. (c) Time average of the
absolute value of $c_i(t)$ in different base states. The existent links are
marked by the blue color, and the nonexistent ones by red, where the threshold
for determining the existent links is $10\%$ of the maximum value.}
\label{fig:S1}
\end{figure}

The time evolution of $\psi_0$ is governed by the Schr\"{o}dinger equation.
The expectation value of operator $A$ is given by
$\langle \psi_t | A|\psi_t\rangle$. In the Heisenberg picture, the states
do not change but the operators change with time. We expand the operator
$A^H(t)$ in the basis $|1\rangle$ and $|0\rangle$, with the corresponding
matrix $\mathbb{A}^H(t)$. The Hermitian property of $\mathbb{A}^H$ stipulates
it must contain $4$ independent elements
\begin{equation}
\mathbb{A}^H(t)=\begin{pmatrix}
A^H_1(t) & A^H_2(t)+iA^H_3(t) \\ A^H_2(t) - iA^H_3(t) & A^H_4(t)
\end{pmatrix}.
\end{equation}
Both the Schr\"{o}dinger and Heisenberg pictures should give the same physical
results. We have
\begin{widetext}
\begin{equation}
(\phi_{11}-i\phi_{12}, \phi_{21}-i\phi_{22})\begin{pmatrix}
A^H_1(t) & A^H_2(t)+iA^H_3(t) \\ A^H_2(t) - iA^H_3(t) & A^H_4(t)
\end{pmatrix}
\binom{\phi_{11}+i\phi_{12}}{\phi_{21}+i\phi_{22}}=\langle \psi_t | A|\psi_t\rangle.
\end{equation}
The unknown elements $A^H(t)$ appear in the equation in a linear fashion:
\begin{equation}
A_1^H(t)(\phi_{11}^2+\phi_{12}^2)+A_2^H(t)(2\phi_{11}\phi_{21}+2\phi_{12}\phi_{22})+A_3^H(t)(2\phi_{12}\phi_{21}-2\phi_{11}\phi_{22})+A_4^H(t)(\phi_{21}^2+\phi_{22}^2)=\langle \psi_t | A|\psi_t\rangle.
\end{equation}
\end{widetext}
At least four different initial conditions are required to solve this equation,
and a further increase in the number of states changes little the result. We
take $A$ to be a Pauli matrix. It expectation value versus time for a given
initial state is illustrated in in Fig.~\ref{fig:S1}(a).

\paragraph*{Step 2:} Let the Hamiltonian of the unknown system be $H(t)$.
The corresponding operator in the Heisenberg picture is $H^H(t)$. Expanding
the operator in the basis, we get
\begin{equation}
\mathbb{H}^H(t)=\begin{pmatrix}
H^H_1(t) & H^H_2(t)+iH^H_3(t) \\ H^H_2(t)-iH^H_3(t) & H^H_4(t)
\end{pmatrix}.
\end{equation}
The quantum evolution is governed by the Heisenberg equation:
\begin{equation}
\frac{dA^H(t)}{dt}=i[H^H(t),A^H(t)].
\end{equation}
For a given initial state, the Heisenberg equation can be written in the
matrix form as
\begin{equation} \label{eq:HE_one_spin}
\frac{d}{dt}\langle A\rangle_t=i\psi_0^\dagger \left[\mathbb{H}^H(t)\mathbb{A}^H(t)-\mathbb{A}^H(t)\mathbb{H}^H(t)\right]\psi_0.
\end{equation}
Expanding the right side, we get a summation of $24$ terms. As the system size
is increased, the number in the summation grows quickly. Through the matrix
product, we get
\begin{equation}
\frac{d}{dt}\langle A\rangle_t=\mathbb{T}\bm{\phi}\bm{\phi}\mathbf{A}(t)\mathbf{H}(t),
\end{equation}
where
\begin{eqnarray}
\nonumber
\bm{\phi} & = & [\phi_{11},\phi_{12},\phi_{21},\phi_{22}]^T, \\ \nonumber
\mathbf{A}(t) & = & [A^H_1(t),A^H_2(t),A^H_3(t),A^H_4(t)]^T, \\ \nonumber
\mathbf{H}(t) & = & [H^H_1(t),H^H_2(t),H^H_3(t),H^H_4(t)]^T,
\end{eqnarray}
and $\mathbb{T}$ is tensor of rank four (with dimension 
$4\times 4\times 4\times4$), which depends only on the dimension of the 
system and is defined as
\begin{equation}
\mathbb{T}\bm{\phi}\bm{\phi}\mathbf{A}(t)\mathbf{H}(t)=\sum_{i,j,m,n}\mathbb{T}_{ijmn}\bm{\phi}_n\bm{\phi}_mA_j(t)H_i(t).
\end{equation}

The left side of Eq.~(\ref{eq:HE_one_spin}) contains the derivatives of the
measurements, which can be determined from the observations. On the right side,
$\mathbf{A}(t)$ and $\bm{\phi}$ are known, so the unknown quantity is
$\mathbf{H}(t)$.

In the HENN, we set the input dimension as one and the output is
$\mathbf{H}(t)$. We choose the batch size to be the number of measurement
points times the number of different initial states. The loss function is
\begin{equation}
\mathcal{L}=\sum_{A=\sigma_x, \sigma_y, \sigma_z}\left| \langle \dot{A}(t) \rangle_\text{real}- \mathbb{T}\bm{\phi}\bm{\phi}\mathbf{A}(t)\mathbf{H}(t) \right|^2.
\end{equation}
We build the neural network from the Keras Tensorflow
package~\cite{chollet2015keras}, where the input is connected to two dense
layers. Constructing this customized loss function is equivalent to
designating a loss function with different weights.

\paragraph*{Step 3:} After the HENN is trained, we input a time series from
$0$ to $5$ and predict the Hamiltonian. The prediction is carried out in
the Heisenberg picture, which can be converted into the corresponding
Hamiltonian in the Schr\"{o}dinger picture. This can be done through an
iteration process.

From the Hamiltonian in the Schr\"{o}dinger picture, we can get the
coefficients in each base through
\begin{equation}
H(t)=c_0(t)\mathbb{I}_2+c_1(t)\sigma_x+c_2(t)\sigma_y+c_3(t)\sigma_z.
\end{equation}
Writing it in a general form $H(t)=\sum_i S_i c_i(t)$, where
$S=\mathbb{I}_2,\sigma_x,\sigma_y,\sigma_z$, we have that $c_i(t)$'s contain
all the information about the Hamiltonian. The predicted $c_i(t)$ in a given
basis are shown Fig.~\ref{fig:S1}(b). The agreement between the solid
(predicted) and dashed (true) curves is proof that the HENN can recover the
Hamiltonian of the original system through observations.

The coupling configuration can be determined through
\begin{equation}
\overline{c}_i=\int |c_i(t)| dt.
\end{equation}
If there exists a coupling between the spin components, the corresponding
coefficient $c_i(t)$ should be non-zero, giving rise to a non-zero value of
$\overline{c}_i$. Figure~\ref{fig:S1}(c) shows the time averaged result of
$c_i(t)$, which has a pronounced value in $\sigma_x$, in agreement with the
original Hamiltonian (\ref{eq:A1_Hamltonian}).

\begin{figure*} [ht!]
\centering
\includegraphics[width=\linewidth]{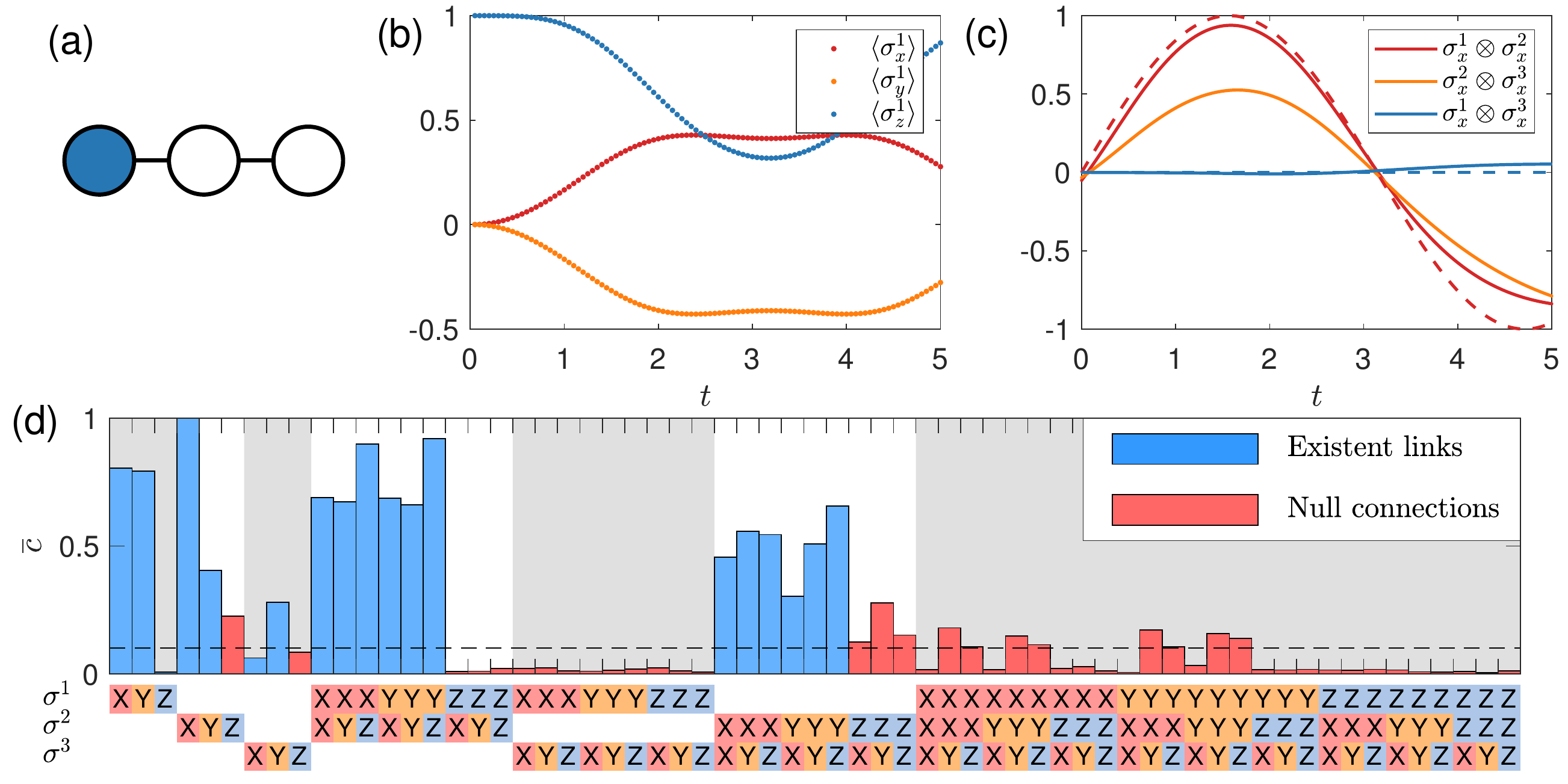}
\caption{ HENN based tomography of a three-spin chain system. (a) Schematic
illustration of the chain, where measurements are taken from the left
spin. (b) Time series of $\sigma^1_x$, $\sigma^1_y$ and $\sigma^1_z$ from the
initial state $\frac{|1\rangle+i|0\rangle}{\sqrt{2}}|11\rangle$, where the
dots correspond to the sampled measurement time series. (c) Predicted
Hamiltonian in the Schr\"{o}dinger picture after decomposition into different
interaction terms, where $\sigma_x^1\sigma_x^2$ and $\sigma_x^1\sigma_x^3$ are
the couplings between the observed node and the hidden nodes, which agree with
the true Hamiltonian terms (dashed traces). The term $\sigma_x^2\sigma_x^3$
specifies the coupling between the two hidden nodes, and the predicted
Hamiltonian is not similar to the true function $\sin(t)$ but not zero
either, so the HENN does predict the existence of this interaction.
(d) Time average of the absolute value of the coefficients $c_i(t)$ associated
with different interaction terms, where the existent links are marked blue
and the nonexistent ones red. The threshold for distinguishing the existent
from nonexistent links is set to be $10\%$ of the maximum coefficient value,
as indicated by the horizontal dashed line.}
\label{fig:S2}
\end{figure*}

\subsection{A three-spin chain}

We consider three interacting spins on a chain, with measurements taken from
the first spin, as shown in Fig.~\ref{fig:S2}. The Hamiltonian is
\begin{equation} \label{eq:A2_chain}
H(t)=\sin(t)\left(\sum_{i=1}^3\sum_{j=1}^2 \sigma_j^i+\sum_{i=1}^2\sum_{l=1}^3\sum_{m=1}^2\sigma_l^i\sigma_m^{i+1} \right).
\end{equation}
There are couplings between spins one and two, and between spins two and
three. For each coupling, there are six links.

The HENN can be constructed following the three steps.

\paragraph*{Step 1:} The system contains $2^3$ independent states:
$|111\rangle, |110\rangle, |011\rangle \cdots |000\rangle$. The initial
conditions are chosen according to Eq.~\eqref{eq:A1_initial}, with the
difference that here there are eight dimensions. Since observations are
taken from the first spin, we write $\sigma_x^1=\sigma_+^1+\sigma_-^1$,
where the creation and annihilation operators act only on the first spin.
The matrix expression for $\sigma_x^1$ is
\begin{displaymath}
\sigma_x^1=\sigma_x\otimes\mathbb{I}_4.
\end{displaymath}
Similarly, we can get the matrices for $\sigma_y^1$ and $\sigma_z^1$. For a
given initial state, we calculate the expectation values of the three
observables on the first spin, as shown in Fig.~\ref{fig:S2}(b).

\paragraph*{Step 2:} Similar to the one-spin system, the Heisenberg equation is
\begin{equation}
\frac{d}{dt}\langle A\rangle_t=\mathbb{T}\bm{\phi}\bm{\phi}\mathbf{A}(t)\mathbf{H}(t),
\end{equation}
where $\bm{\phi}$ is a vector of $2\times 2^3=16$ elements, $\mathbf{A}(t)$
and $\mathbf{H}(t)$ contain $4^3=64$ elements, and $\mathbb{T}$ is a tensor
of rank four with the dimension $64 \times 64 \times 16\times 16$:
\begin{equation}
\mathbb{T}\bm{\phi}\bm{\phi}\mathbf{A}(t)\mathbf{H}(t)=\sum_{i,j,m,n}\mathbb{T}_{ijmn}\bm{\phi}_n\bm{\phi}_mA_j(t)H_i(t).
\end{equation}

We build up the HENN according to the same loss function as in the case of
a single-spin system, predict the Hamiltonian, and convert it to the
Schr\"{o}dinger picture. The Hamiltonian can be decomposed as
\begin{equation}
\begin{split}
\mathbb{H}(t)&= c_0(t)\mathbb{I}+\sum_{i,j} c_{i,j}(t) \sigma^i_j\\
&+ \sum_{i,j,m,n} c_{ijmn}(t) \sigma^i_j \sigma^m_n\\
&+\sum_{i,j,m,n,k,l} c_{ijmnkl}(t) \sigma^i_j \sigma^m_n\sigma^k_l.\\
\end{split}
\end{equation}
The decomposition becomes cumbersome for systems with more than one
spin. we thus write this as the direct product of the Pauli matrices plus
the identity matrix. For example, the two-body coupling $\sigma_1^1\sigma_2^2$
can be written as $\sigma_1^1\otimes\sigma_2^2\otimes\sigma_0^3$.

Figure~\ref{fig:S2}(c) shows the predicted Hamiltonian in several base states.
The Hamiltonian for the coupling $\sigma_x^1\sigma_x^2$ can be compared with
the sinusoidal function $\sin(t)$. The coupling term $\sigma_x^1\sigma_x^3$
is non-existent in the original system, so it should be compared with zero.
The agreement indicates that the local Hamiltonian between the observed spin
and the hidden spins can be recovered. Note that $\sigma_x^2\sigma_x^3$
represents a coupling between the two hidden nodes, whose true value is
$\sin(t)$, but the predicted Hamiltonian is not close to it. Nonetheless,
the non-zero value of the predicted term indicates the existence of the
coupling term $\sigma_x^2\sigma_x^3$.

\paragraph*{Step 3:} After decomposing the Hamiltonian in different terms,
we take the time average of each and normalize them, as shown in
Fig.~\ref{fig:S2}(d). The ideal case is that all the blue points have the
value one and all the red points are zero. First, the Hamiltonian
Eq.~\eqref{eq:A2_chain} contains self-couplings in $x$ and $y$ but not in $z$,
which are indicated by the first three bars in Fig.~\ref{fig:S2}(d). For
the hidden spins, there are errors in predicting the self-coupling terms.
Second, there are two-body interactions between spins one and two, and between
spin two and three, but not between one and three, where each existent
interaction has six terms of coupling. The predicted results for the couplings
involving the first spin are more accurate than those between the hidden spins.
Third, the true Hamiltonian does not include any three-body interactions, so
all such terms should be zero.

The results in Fig.~\ref{fig:S2} indicate that our HENN can perform accurate
tomography of the three spin chain.

\section{Time-independent Toffoli and Fredkin gates} \label{Appendix_B}

The Hamiltonian for the time-independent Toffoli gate is
\begin{equation}
H_\text{Toffoli}=\frac{\pi}{8} (\mathbb{I}_2-\sigma^1_3)(\mathbb{I}_2-\sigma^2_3)(\mathbb{I}_2-\sigma^3_1).
\end{equation}
The corresponding time evolution operator is
\begin{equation}
U_\text{Toffoli}=\begin{pmatrix}
\mathbb{I}_6 & \\
& \mathbb{X}_0(t)
\end{pmatrix},
\end{equation}
where
\begin{equation}
\mathbb{X}_0(t) =\frac{1}{2}\begin{pmatrix}
1+\exp(i\pi t) & 1-\exp(i\pi t) \\
1-\exp(i\pi t) & 1+\exp(i\pi t)
\end{pmatrix}.
\end{equation}
The Hamiltonian for the time-independent Fredkin gate is
\begin{equation}
H_\text{Fredkin}=\frac{\pi}{8} (\mathbb{I}_2-\sigma_1)\left[\mathbb{I}_4-\sum_{\alpha=1}^3 \sigma_2^\alpha\sigma_3^\alpha \right],
\end{equation}
with
\begin{equation}
U_\text{Fredkin}=\begin{pmatrix}
\mathbb{I}_5 & & \\
& \mathbb{X}_0(t) & \\
& & 1
\end{pmatrix}.
\end{equation}


%
\end{document}